# A Secure Two-Party Computation Protocol for Intersection Detection between Two Convex Hulls

Amirhmad Chapnevis[1], Babak Sadeghiyan[2]

[1]Department of Computer Engineering and Information Technology, Amirkabir University of Technology, Tehran, Iran
chapnevis@aut.ac.ir

[2]Department of Computer Engineering and Information Technology, Amirkabir University of Technology, Tehran, Iran
basadegh@aut.ac.ir

**Abstract**

Intersection detection between three-dimensional bodies has various applications in computer graphics, video game development, robotics as well as military industries. In some respects, entities do not want to disclose sensitive information about themselves, including their location. In this paper, we present a secure two-party protocol to determine the existence of an intersection between entities.

The protocol presented in this paper allows for intersection detection in three-dimensional spaces in geometry. Our approach is to use an intersecting plane between two spaces to determine their separation or intersection. For this purpose, we introduce a computational geometry protocol to determine the existence of an intersecting plane. In this paper, we first use the Minkowski difference to reduce the two-space problem into one-space. Then, the separating set is obtained and the separation of two shapes is determined based on the inclusion of the center point. We then secure the protocol by modifying the separating set computation method as a privacy-preserver and changing the Minkowski difference method to achieve this goal. The proposed protocol applies to any form of convex three-dimensional shape. The experiments successfully found a secure protocol for intersection detection between two convex hulls in geometrical shapes such as the pyramid and cuboid.

## 1- Introduction

Computational geometry is a branch of computer science that studies geometrical algorithms. In general, computational geometry is important for solving problems whose solutions require extensive operations that can be substituted with geometric shapes. Furthermore, several geometrical problems require computational geometry algorithms for a solution [1] In this respect, the Art Gallery problem can serve as an example where a polygon is considered as a gallery, and the problem is to find the minimum number of cameras needed to monitor the whole place for protection.

Another major use of computational geometry is for problems in computer graphics, robots, and artificial intelligence [2] For example, in video games, computational geometry algorithms are used to determine whether or not two or three-dimensional shapes collide with each other. In the aviation industry, other geometric algorithms are used to find the distance between two aircraft.

In these problems, the privacy-preservation of entities can be especially important. Privacy preservation means the sensitive data of entities partaking in geometrical algorithms should not be revealed to other parties. Also, after the algorithm is executed and each entity has reached an answer, with the exception of intermediate data transferred during the execution and the final answer, the participants should not gain access to others' sensitive data (directly or through computation).

In terms of information security, there is a set of protocols called secure multi-party computation whose main purpose is preserving the privacy of each entity during the algorithm's execution [3]. Secure multi-party problems aimed at preserving privacy are divided into five categories according to their applications. One of these groups is the secure multiparty protocols for computational geometry problems.

Computational geometry protocols with privacy preservation are secure multi-party protocols that allow the parties to use them to obtain a certain output while ensuring the privacy of their data. These protocols were first implemented in 2001 by Atallah and Wenliang Du [4] by providing higher-performance solutions compared to circuit evaluation protocols. These protocols are used in various applications such as wireless sensor networks in the military as well as location finders in vehicles. In both these applications, preserving the privacy of each entities' data is of particular importance. In the military wireless sensor systems, the coordinates of each entity are critical and the privacy of data that can reveal the sensors' location must be preserved. Moreover, in the car location finders, since the vehicle's location is sent to the GPS various applications, privacy-preservation is essential to users. Thus, the need for privacy-preserving protocols in computational geometry is growing every day.

The privacy-preserving computational geometry protocols must be fundamentally applied to geometrical shapes, and their properties, strengths, and weaknesses evaluated before they are used for such applications. There have been limited studies on privacy-preserving protocols, and in some cases, the results do not work accurately in three-dimensional spaces [3,6]. Other drawbacks of the protocols proposed in this area are their limitation to two-dimensional space. Meanwhile, in computational geometry, there are various algorithms for determining the different modes of planes and other geometrical shapes in three-dimensional spaces [6]. Note that in reality, shapes exist as different and undefined geometrical forms. For example, three-dimensional shapes should be examined to study the vehicles. Therefore, to study the privacy preservation protocols in three-dimensional space, it is necessary to analyze undefined shapes such as convex hulls formed from a series of points in space.

Based on the literature, studies on privacy-preserving computational geometry protocols for three-dimensional shapes are limited. The research in this field is limited to specific forms, such as pyramids, cones, and several other three-dimensional shapes [6]. Also, in some articles, the results obtained for geometrical shapes are expressed in approximations. We assume there are two convex hulls, each belonging to one entity. Each convex hull consists of the smallest convex form that can cover all the points. This paper is aimed at finding a protocol for determining the existence of an intersection between two convex hulls in three-dimensional space with an acceptable performance - i.e., preserving the privacy of each entity. In obtaining the result on the existence or absence of an intersection between the two convex hulls, no other information —such as the properties of the points and the convex hull of each entity— should be disclosed.

The arrangement of the content in this paper is as follows: Several intersection detection algorithms in computational geometry are presented briefly in the second section. The algorithm used in the proposed protocol along with its prerequisites is presented as well. Furthermore, the privacy-preserving scalar product protocol used as the solution in the proposed protocol is presented in detail. In the third section, the proposed protocol is presented. Changes in the geometric algorithm and the implementation of the privacy-preserving scalar product protocol are included in this chapter. Finally, in the fourth chapter, results and conclusions are presented.

## 2- Background
In this section, several computational geometry algorithms proposed for intersection detection regardless of privacy are discussed. Also, the privacy-preserving scalar product protocol is presented as a widely used solution in privacy-preserving computational geometry problems. This approach has a higher performance and a lower time complexity compared to other protocols.

## 2-1- Computational geometry algorithm for intersection detection between convex hulls using a sphere [2]
The algorithm presented in this section is an optimized solution to the problem. If one of the convex hulls has $P$ planes and the other one has $Q$ planes, the time complexity for determining the existence of intersection between them is equal to $O(log|P| + log|Q|)$. In this algorithm, Minkowski difference can be used to simplify the execution and study only one shape. In this method, the two shapes $A$ and $B$, partaking in the geometric algorithm for intersection detection, are transformed into a single shape. Using the Minkowski difference to Subtract shape $A$ from shape $B$ results in a new shape that can be used for intersection detection computation. If the final shape contains the point of origin, it indicates the existence of an intersection between the shapes $A$ and $B$. In Figure 1, this claim is presented:

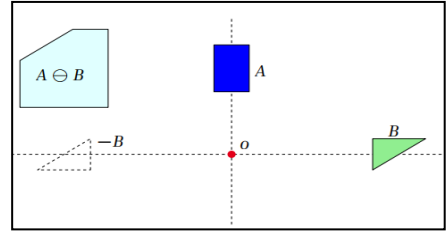

**Fig. 1. An examination of one shape instead of two shapes using Minkowski difference**

$A \ominus B = \{a - b | a \in A, b \in B\} = A \oplus (-B);$ (1)

$(A \cap B) = \phi \Leftrightarrow o \notin (A \ominus B);$ (2)

**Support function**
The support function for a convex hull (non-empty) is defined as:

$h_A(x) = \sup(x.a: a \in A);$ (3)

It is equal to the maximum internal multiplication value of the unit vector x by vector a, so the point $a$ belongs to the convex hull.

The support function is
$h_p$ is: $S^2 \to R, n \to \max_{x \in P}(x.n);$ (4)

**Extremal function**
The Extremal function is defined using the concept of support function: This function's output is equal to the point in the convex hull in the direction n where the support function is at its highest. In other words, the extremal function is defined as:

The extremal function is $\sum_P : S^2 \to R^3, n$
$\to \text{argmax}_{x \in P}(x.n);$ (5)

If the point of origin is not within the convex hull shape, there must, therefore, be a point that belongs to the convex hull and a direction based on which the support function is negative. A negative support function means that there must be a line (perpendicular to the vector of direction n) that passes through the point of origin of the coordinate and the convex hull. In Figure 2, as the center point is located outside, the support function has a negative value in part of the shape.

The extremal function is used to simply calculate only the specific points in this shape. Consider the following equation:

$$h_{A \ominus B}(n) = \max_{a \in A}(n.a)$$
$$- \min_{b \in B}(n.b)$$
$$= h_A(n) + h_B(-n); \quad (6)$$
$$\sum_{A \ominus B}(n) = \sum_A(n) - \sum_B(-n); \quad (7)$$

The equation is calculated with the time complexity $O(|A|+|B|)$ (without the need to obtain a final shape from the difference between the two shapes). In Fig. 3, the symbols used in Formula 7 can be understood more accurately.

The main idea of this paper is presented in this section, which allows for determining whether or not the point of origin is located in P (which is the Minkowski difference of shapes) and consequently, whether or not an intersection exists between the two shapes—i.e., the main answer to the problem.

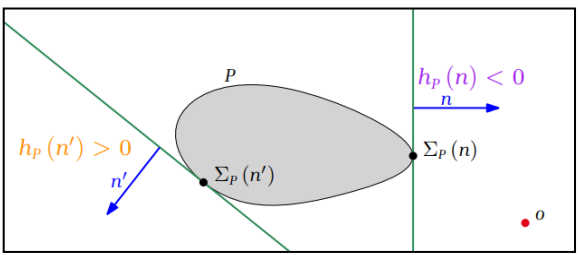

**Fig. 2. The negativity of support function at least at one point which results in the origin of coordinates being outside the shape.**

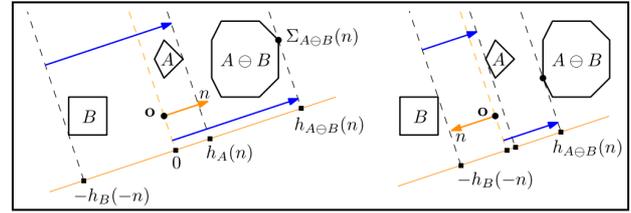

**Fig. 3. Definition of the support function, the extremal function, and the Minkowski difference in the shape**

**Separating set**

The separating set of the point P for the convex hull is the sum of all points in the geometrical shape where the support function for the tangent line from that point is negative (the direction perpendicular to it is considered the vector). Each point of the shape is known as a vector of origin of the coordinates.

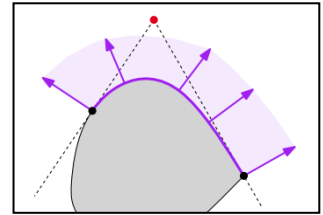

**Fig. 4. Separating set of a shape**

defining separating set x
$$= \{ n.x < 0 \}; \quad (8)$$

**Algorithm for determining the existence of an intersection between two geometrical shapes**

1- Arbitrary choice of a direction based on the separating set (specifying the direction of n)
2- The determination of the point is based on the extremal function according to the direction of n.

$$v_{i+1} = \sum_P(n) \quad (9)$$

3- Obtaining the support function for the specified point and the specified direction. If the support function value is negative, the algorithm ends and it is concluded that the origin of the coordinates is not in the shape, and therefore, the two shapes do not have an intersection.

$$if\ h_P(n) < 0 \Rightarrow o \notin P; \tag{10}$$

4- If the support function value is positive, it is necessary to obtain a more accurate estimation based on the separating set, and the function is called once again. This should be continued until the separating set is empty to conclude that the origin of the coordinates is within the shape P, and consequently, the two shapes have an intersection. The best approximation is to choose the median of the separating set in each step.

$$\text{If } o \in P_{i+1} \Rightarrow o \in P \tag{11}$$

In addition to the proposed algorithm, there are others for intersection detection in computational geometry, such as V-CLIP [7] or GJK algorithms [8]. The V-CLIP algorithm uses Voronoi diagram methods and its principles to determine the direction. The GJK algorithm method determines the distance between two convex hulls in addition to the existence of an intersection between them. Another algorithm worth mentioning is Kirkpatrick's hierarchical algorithm [9], which was not considered due to its time complexity.

## 2.2 The privacy-preserving scalar product protocol [10].

In this protocol, each entity is assumed to have a vector, and the vector is sensitive data whose privacy must be preserved. By using the privacy-preserving scalar product protocol, each entity can obtain the scalar product value of the two vectors without the disclosure of any other information on the other party's vector. Various protocols have been proposed for this purpose, and the one with the best time complexity is presented here. In the implementation process, the protocol uses the homomorphic encryption schemes, as well as the vector summation protocol. The following relation is held at an encryption system, provided it is capable of summation homomorphism:

$$E(x_1) \times E(x_2) = E(x_1 + x_2) \tag{12}$$

**Vector addition protocol**

Suppose, Alice has a vector named $x$. Bob has a vector called $y$, as well as a permutation called π. The goal after the implementation of the protocol is that Alice can obtain the value of $\pi(x + y)$. For this purpose, Alice produces a key pair for an asymmetric homomorphic encryption system. The key pair consists of a public key and a private key exclusive to her. Then she sends the public key to Bob. Alice encrypts her exclusive vector $x$ one entry at a time and sends it to Bob.

$$x = (x_1, \dots, x_n)^T; \tag{13}$$

$$E(x) = \big(E(x_1), \dots, E(x_n)\big)^T; \tag{14}$$

Bob, using Alice's public key, encrypts vector $y$ one entry at a time, and then using the addition feature in the homomorphic encryption system, adds it to the first encrypted vector. At last, the permutations are obtained on all the entries. This final vector, which is $\pi(E(x + y))$, is sent to the Alice.

$$E(y) = \big(E(y_1), \dots, E(y_n)\big); \tag{15}$$

$$E(x + y) = E(x) \times E(y); \tag{16}$$

**Privacy-preserving scalar product protocol**

In this protocol, Alice and Bob each have a private individual vector, doubled a and b, respectively. each vector consists of n entries. After the implementation of the protocol, Alice and Bob need to obtain the value of $a.b$ . Consider the following equations:

$$2a_ib_i = a_i^2 + b_i^2 - (a_i - b_i)^2; \tag{17}$$

$$2\sum_{i=1}^{n} a_ib_i = \sum_{i=1}^{n} a_i^2 + \sum_{i=1}^{n} b_i^2 - \sum_{i=1}^{n}(a_i - b_i)^2; \tag{18}$$

**Protocol procedure:**

1- Alice and Bob use the privacy-preserving vector addition protocol. Alice obtains the value of $\pi_0(a - b)$, where $\pi_0$ is the permutation determined by Bob.

2- Alice computes the following value:

$$\sum_{\pi_0(i)=1}^{n} \left(a_{\pi_0(i)} - b_{\pi_0(i)}\right)^2 + \sum_{i=1}^{n} a_i^2; \qquad (19)$$

3- Bob also calculates the $\sum_{i=1}^{n} b_i^2$. Then, Bob sends the calculated value to Alice. Alice, using the value received, will be able to obtain the scalar product. Note that during the protocol implementation, the privacy of each entity is protected.

$$2a.b = 2\sum_{i=1}^{n} a_i b_i$$

$$= \sum_{i=1}^{n} a_i^2$$

$$+ \sum_{i=1}^{n} b_i^2$$

$$- \sum_{\pi_0(i)=1}^{n} \left(a_{\pi_0(i)} - b_{\pi_0(i)}\right)^2; (20)$$

### 3. The proposed protocol

The present protocol is presented through the modification and composition of existing computational geometry algorithms used for intersection detection between two convex hulls in three-dimensional space and based on data securing methods for preserving privacy. Note that the proposed protocol is not only used for intersection detection in three-dimensional space but also improves the privacy-preserving intersection detection protocols between two polyhedrons that have been presented via various methods so far.

Before expressing the proposed protocol, it is necessary to discuss the implementation of the algorithm. In the proposed algorithm, the two entities, each having convex hull properties, need to provide information on their shapes to the other party in only one step of the process. This step is seen from the very beginning of the algorithm which requires the calculation of Minkowski difference. Per the algorithm, a Minkowski subtraction of the two points —each belonging to one entity— is required with each function call. With a closer examination of the algorithm, we found the accurate computation of the vector is not necessary for calculating the difference of the two vectors, and that it is sufficient to calculate the difference vector's direction. The reason behind this is better understood by viewing the Figure 5.

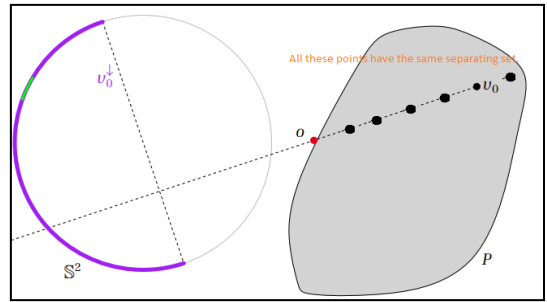

**Fig. 5. Same separating set at all points in one direction**

According to the figure 5, the main reason why this happens lies in the computation of the separating set. The separating set was calculated for the point $v_0$ (with respect to the origin of coordinates). It is to be noted that if this point was located elsewhere on the line between itself and the origin, the separating set would still be the same. Also, this point was obtained through the subtraction of two points from each shape. Therefore, it is sufficient to provide a secure multi-party protocol to calculate the direction of difference between the two vectors to obtain the objective of the problem — i.e., privacy-preserving intersection detection between the two geometrical shapes — by combining the proposed protocol and the geometrical algorithm.

By having each of the following set of information, a vector can be identified accurately:

- Having the minimum values of x, y, and z.
- Having the minimum value of r and angles θ and φ.

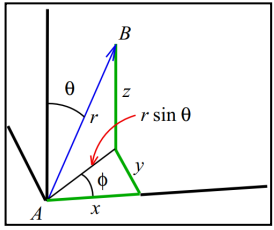

**Fig. 6. Angles forming a vector in space**

Therefore, to keep each entity from accurately discovering the vector for privacy-preserving, information disclosure on the sum vector should be limited. By having angles and, the direction of the sum vector can be calculated accurately. However, since the value $r$ remains confidential, entities cannot identify its exact value and locate the vector precisely. According to the figure 6, the angle values can be calculated using the relations provided here:

$$\theta = \tan^{-1}(\frac{\sqrt{x^2 + y^2}}{z}); \qquad (21)$$

$$\phi = \tan^{-1}\left(\frac{y}{x}\right); \qquad (22)$$

Since the vector angles obtained are only for a given vector, it should be noted that in the geometrical algorithm the sum vector is made of the vectors X and Y, each belonging to one entity. Therefore, each entry should be written as the sum of the two vector entries for X and Y. We assume:

The vector X of entity A and the vector Y of entity B are represented as $X = (x_1, y_1, z_1)$ and $Y = (x_2, y_2, z_2)$, respectively. Therefore, $\phi$ and $\theta$ are the constituent angles of vector $X + Y$ that must be provided with privacy-preservation to each entity. These angles are generated by the sum of vectors $X$ and $Y$ entries. Therefore:

$$\theta_{X+Y} = \tan^{-1}(\frac{\sqrt{(x_1 + x_2)^2 + (y_1 + y_2)^2}}{(z_1 + z_2)}) \quad (23)$$

And in the same way:

$$\phi_{X+Y} = \tan^{-1}\left(\frac{y_1 + y_2}{x_1 + x_2}\right); \qquad (24)$$

To use the scalar product protocol for privacy-preserving, it is first necessary for each entity to generate data to introduce to the protocol so that after the computation, the objective of the problem —i.e., angles of the sum vector— is obtained. Changes are made to the target value of the protocol execution to make understand the protocol easier. In changing the protocol output, instead of calculating the angles $\phi$ and $\theta$, the following values are computed so that the process is simplified and the radical operator is eliminated from the final value.

Later, we will explain how the sign can be detected by applying a radical operation:

$$\phi_{X+Y} = \tan^{-1}(\frac{y_1 + y_2}{x_1 + x_2})$$
$$\Rightarrow \text{Protocol objective:} \left(\frac{y_1 + y_2}{x_1 + x_2}\right)^2 ; \qquad (25)$$

$$\theta_{X+Y} = \tan^{-1}(\frac{\sqrt{(x_1 + x_2)^2 + (y_1 + y_2)^2}}{(z_1 + z_2)})$$
$$\Rightarrow \text{protocol objective:} \frac{(x_1 + x_2)^2 + (y_1 + y_2)^2}{(z_1 + z_2)^2}$$
$$= \frac{(x_1 + x_2)^2}{(z_1 + z_2)^2} + \frac{(y_1 + y_2)^2}{(z_1 + z_2)^2}; \qquad (26)$$

In this section, the new protocol is presented by modifying the privacy-preserving scalar product protocol. Each entity is required to compute the protocol's expected values and provide it to privacy-preserving scalar product protocol.

**Protocol assumptions**

The entity $A$ has an $X$ vector, its entries' values are considered private data, and is defined as:

Vector X, Entity A: $(x_1, y_1, z_1)$;

In the same way, entity $B$ has the $Y$ vector, its entries' values are considered private data, and is defined as:

Vector Y, Entity B: $(x_2, y_2, z_2)$;

According to the protocol assumptions, each entity must compute the data required by the protocol and provide them to the privacy-preserving scalar product protocol.

In addition to the explanations, it is necessary to note that the proposed protocol works with the assumption that the entities are semi-honest. Being semi-honest means that entities must act based on the procedure presented in the protocol, but they will be able to use protocol output as well as intermediate data obtained in the implementation process to perform any arbitrary computations.

**Protocol output**
In the output, the target values must be at the disposal of each of the entities without disclosing any other information about the entities' sensitive data. To make the protocol simpler, the process' objective in the first stage is to obtain this value:

$$\phi_{X+Y} = \tan^{-1}\frac{y_1 + y_2}{x_1 + x_2}$$
$$\Rightarrow \text{Protocol objective:} \left(\frac{y_1 + y_2}{x_1 + x_2}\right)^2; \qquad (27)$$

**Protocol procedure**
**Step one:** The computation should be carried out privately by each entity.

Entity A: $\left(\dfrac{x_1^2}{y_2^2 z_1^2}, \dfrac{1}{y_2^2 z_1^2}, \dfrac{2x_1}{y_1^2 z_1^2}\right)$; (28)

Entity B: $\left(\dfrac{1}{y_2^2 z_2^2}, \dfrac{x_2^2}{y_2^2 z_2^2}, \dfrac{x_2}{y_2^2 z_2^2}\right)$; (29)

Each entity computes the presented values and participates in the privacy-preserving scalar product protocol. The protocol's output, which is obtained in the first step, is equivalent to:

$$\left(\frac{x_1^2}{y_1^2 z_1^2}.\frac{1}{y_2^2 z_2^2}\right) + \left(\frac{1}{y_1^2 z_1^2}.\frac{x_2^2}{y_2^2 z_2^2}\right) + \left(\frac{2x_1}{y_1^2 z_1^2}.\frac{x_2}{y_2^2 z_2^2}\right)$$
$$= \frac{(x_1 + x_2)^2}{y_1^2 y_2^2 z_1^2 z_2^2}; \quad (30)$$

Each entity inverts the output received from the privacy-preserving scalar product protocol to use

them in computing the next inputs from the protocol. Thus, we will have:

$$\left(\frac{(x_1 + x_2)^2}{y_1^2 y_2^2 z_1^2 z_2^2}\right)^{-1} = \frac{y_1^2 y_2^2 z_1^2 z_2^2}{(x_1 + x_2)^2}; \qquad (31)$$

Each entity will have access to this value, and it should be noted that the value will not reveal any information about the other entity's vector and sensitive data.

**Step Two:** The protocol continues in the second step as each entity carries out computations similar to the first step. Before entities calculate the values of the privacy-preserving scalar product protocol for the second stage, it is necessary to make changes to the output from the previous step. The point of enumeration in the formula is to clarify that each computed value is later sent to which entry as input for the privacy-preserving scalar product protocol. The changes made by entity A are as follows:

$$\frac{y_1^2 y_2^2 z_1^2 z_2^2}{(x_1 + x_2)^2} \xrightarrow[\text{divided by } y_1 z_1^2]{} \frac{y_1 (y_2^2 z_2^2)}{(x_1 + x_2)^2} \; \mathbf{❶}; \qquad (32)$$

$$\frac{y_1^2 y_2^2 z_1^2 z_2^2}{(x_1 + x_2)^2} \xrightarrow[\text{divided by } y_1^2 z_1^2]{} \frac{y_2^2 z_2^2}{(x_1 + x_2)^2} \; \mathbf{❷}; \qquad (33)$$

Similarly, the changes made by the entity B are:

$$\frac{y_1^2 y_2^2 z_1^2 z_2^2}{(x_1 + x_2)^2} \xrightarrow[\text{divided by } y_2^2 z_2^2]{} \frac{y_1^2 z_1^2}{(x_1 + x_2)^2} \; \mathbf{❸}; \qquad (34)$$

**Step three:** After the computation is carried out by each entity, the values should be provided privately for the privacy-preserving scalar product protocol in the second step, and given again as input to the algorithm:

Entity A
: $\left(\dfrac{y_2^2 z_2^2}{(x_1 + x_2)^2} \mathbf{❷}, \dfrac{1}{z_1^2}, \dfrac{y_1 (y_2^2 z_2^2)}{(x_1 + x_2)^2} \mathbf{❶}\right)$; (35)

Entity B
: $\left(\dfrac{1}{z_2^2}, \dfrac{y_1^2 z_1^2}{(x_1 + x_2)^2} \mathbf{❸}, \dfrac{2}{y_2 z_2^2}\right)$; (36)

After the entities perform the privacy-preserving scalar product protocol, the output will be transferred to the parties as below, and no other sensitive data will be disclosed:

Entity A
$$: \left( \frac{y_2^2 z_2^2}{(x_1 + x_2)^2} ❷, \frac{1}{z_1^2}, \frac{y_1(y_2^2 z_2^2)}{(x_1 + x_2)^2} ❶ \right); \qquad (37)$$

Entity B
$$: \left( \frac{1}{z_2^2}, \frac{y_1^2 z_1^2}{(x_1 + x_2)^2} ❸, \frac{2}{y_2 z_2^2} \right); \qquad (38)$$

The output of the privacy-preserving scalar product protocol in the second step will be as follows:

$$\left( \frac{y_2^2 z_2^2}{(x_1 + x_2)^2} \cdot \frac{1}{z_2^2} \right) + \left( \frac{1}{z_1^2} \cdot \frac{y_1^2 z_1^2}{(x_1 + x_2)^2} \right)$$
$$+ \left( \frac{y_1(y_2^2 z_2^2)}{(x_1 + x_2)^2} \cdot \frac{2}{y_2 z_2^2} \right)$$
$$= \frac{(y_1 + y_2)^2}{(x_1 + x_2)^2}; \qquad (39)$$

According to the computations, entities are made aware of the second output value of the privacy-preserving scalar product protocol. Every entity must calculate the square root of the received value, but since the square root can be positive or negative, another approach — which is presented in the next section— can be used to determine the sign.

$$\tan \phi = \sqrt{\frac{(y_1 + y_2)^2}{(x_1 + x_2)^2}} = \pm \frac{y_1 + y_2}{x_1 + x_2}; \quad (40)$$

A similar procedure is used to obtain the value of $\tan \theta$. Of course, given the difference in the formula for $\tan \theta$, the proposed protocol must be executed again and the outputs assembled.

$$\frac{(x_1 + x_2)^2 + (y_1 + y_2)^2}{(z_1 + z_2)^2}$$
$$= \frac{(x_1 + x_2)^2}{(z_1 + z_2)^2} ①$$
$$+ \frac{(y_1 + y_2)^2}{(z_1 + z_2)^2} ②; \quad (41)$$

1- Using the protocol, the two entities are informed of the value of ① while privacy is preserved.

2- Using the protocol, the two entities are informed of the value of ② while privacy is preserved.

3- The values one and two are added together by each entity and their square root is derived. An approach provided in the next section is used to determine the sign.

**Determining the sign of protocol output value**

**Step four:** Since $\tan \phi$ is equal to $\frac{\sin \phi}{\cos \phi}$, the resulting $\sin \phi$ and $\cos \phi$ signs can be combined to determine the final sign.

**Protocol input**
vectors A and B

**Protocol output**
The sign of the A+B vector tangent with privacy-preserving for each entity.

**Protocol procedure**
1- Each entity regards the vectors as the sum of the unit vectors.

2- They provide the sign of each vector value to each other based on the following table.

3- The value of $\sin(A + B)$ is obtained according to the table.

If the j values were not identical, the size of vectors should be compared to each other using the millionaire's problem protocol, and the sign of the larger vector is given to the $\sin(A + B)$.

**Table 1 Determining the sign of the sum of the two vectors' sinus**

| $j_A$ | $j_B$ | $\sin(A + B)$ |
|-------|-------|---------------|
| + | + | + |
| - | - | - |
| + | - | Using the Millionaires' problem protocol and determining the sign based on the magnitude of each vector |
| - | + | |

Similarly, it is applied to unit vectors $i$ and the sign can also be determined using the table below.

**Table 2 determining the sign of the sum of the two vector's cosines**

| $i_A$ | $i_B$ | $\cos(A + B)$ |
|-------|-------|---------------|
| + | + | + |
| - | - | - |
| + | - | Using the Millionaires' problem protocol and determining the sign based on the magnitude of each vector |
| - | + | |

If the j values were not identical, the size of vectors should be compared to each other using the millionaire's problem protocol, and the sign of the larger vector is given to the $\cos(A + B)$.

4- Finally, after specifying the marks of values $\sin \phi$ and $\cos \phi$, each party can be notified of the value sign of $\tan \phi$.

## 4. Conclusion

In this paper, a secure privacy-preserving multi-party protocol was presented to determine the existence of an intersection between two convex shapes in three-dimensional space. The applications of this protocol include use in the military and robotics.

To present the protocol, first, the different types of algorithms available in computational geometry were studied. Also, methods used in secure multi-party protocols for privacy-preserving were examined. Among secure multi-party protocols in computational geometry, the use of the privacy-preserving scalar product, as well as the millionaires' problem, was examined as widely used protocols.

To solve the problem, the conventional privacy-preserving scalar product protocol was defined with new inputs. Thus, we have been able to carry out the Minkowski addition (subtraction) while preserving privacy. Using the proposed protocol, entities can obtain the direction of the sum vector of two vectors while protecting privacy. Since this operation is used in the intersection detection algorithms between entities in computational geometry, by changing the algorithm and using the privacy-preserving scalar product protocol, the privacy of the entities' sensitive data (in this case, points in the convex hulls) is preserved, and at the same time intersection detection is possible for both parties. The main idea in this paper is to eliminate a part of the proposed geometric algorithm, as well as changing the input of the privacy-preserving scalar product protocol.

Besides carrying out the Minkowski addition, only the sum vector angles were computed in the protocol and since its magnitude does not matter in the geometrical algorithm, it is not calculated at the protocol. Therefore, entities can only have access to the direction of the sum vector and not its size. In the future, protocols can be presented for moving objects so results can be obtained before the possible collision.